# On the supremum of the steepness parameter in self-adjusting steepness based schemes

Yucang Ruan[a,b], Baolin Tian[c,d], Xinting Zhang[b,*], Zhiwei He[c,*]

[a] *State Key Laboratory for Turbulence and Complex Systems, College of Engineering, Peking University, Beijing 100871, China*

[b] *Sino-French Engineer School, Beihang University, Beijing 100191, China*

[c] *Institute of Applied Physics and Computational Mathematics, Beijing 100094, China*

[d] *HEDPS and Center for Applied Physics and Technology, College of Engineering, Peking University, Beijing 100871, China*

**Abstract**

Self-adjusting steepness (SAS)-based schemes [1] preserve various structures in the compressible flows. These schemes provide a range of desired behaviors depending on the steepness-adjustable limiters with the steepness measured by a steepness parameter. These properties include either high-order properties with exact steepness parameter values that are theoretically determined or having anti-diffusive/compression properties with a larger steepness parameter. Nevertheless, the supremum of the steepness parameter has not been determined theoretically yet. In this study, we derive a universal method to determine the supremum using Sweby's total variation diminishing (TVD) condition [2]. Two typical steepness-adjustable limiters are analyzed in detail including the tangent of hyperbola for interface capturing (THINC) limiter and the steepness-adjustable harmonic (SAH) limiter. We also obtain the analytical expression of the supremum of the steepness parameter which is dependent on the Courant-Friedrichs-Lewy (CFL) number. Using this solution, we then propose supremum-determined SAS schemes. These schemes are further extended to solve the compressible Euler equations.

*Corresponding authors.

Email addresses: xtzhang@buaa.edu.cn (Xinting Zhang), he_zhiwei@iapcm.ac.cn (Zhiwei He)

The results of typical numerical tests confirm our theoretical conclusions and show that the final schemes are capable of sharply capturing contact discontinuities and minimizing numerical oscillations.

THINC, contact discontinuity, anti-diffusion, total-variation-diminishing, CFL number

# Introduction

Compressible flows including shocks and contact discontinuities are widely observed in scientific research and engineering applications. Rayleigh-Taylor instability [3,4] and Richtmyer-Meshkov [5,6] instability are typical examples of such compressible flows. The evolution of such flows also significantly influences the ignition process of inertial confinement fusion (ICF) [7–9]. Therefore, accurate prediction of various discontinuities in such flows is of great importance. Classical upwind schemes such as the piecewise constant reconstruction Godunov scheme [10], monotonic upstream schemes for conservation laws (MUSCL) [11,12], essentially non-oscillatory (ENO) [13–15], and weighted essentially non-oscillatory (WENO) schemes [16,17], obtain the effective capacity for the capturing shocks. These schemes have an intrinsic advantage in terms of their numerical stability. In other words, if applied to capturing the contact discontinuities, they may encounter large numerical dissipations which inevitably result in smearing the contact discontinuity/material interface. Also, large numerical dissipations are reported in the previous works in cases where such schemes are applied to the linear advection equation after a long period [18].

In contrast, downwind schemes have been rarely used due to their numerical instability. For instance, Després and Lagoutière developed the limited downwind scheme [19] with improved numerical stability by limiting the downwind schemes with a Sweby's TVD conditions [2]. The limited downwind scheme also preserves sharp discontinuities and minimizes numerical oscillations. Their work was extended further by Xu and Shu for anti-diffusive flux corrections [20]. Such schemes are however unable to capture smooth structures because they convert a Gaussian function to a step-like function. To address this issue, Xiao [21] proposed the use of a hyperbolic tangent function to simulate the discontinuities in the grid cells and designed the tangent of hyperbola for interface capturing (THINC) scheme. To control the thickness of the discontinuity, they also introduced an undetermined maneuverable parameter. Although this scheme can be employed in boundary variation diminishing (BVD) schemes[18,22,23], determining the adjustable parameter remains an open problem.

Recently, He [1] demonstrated that the THINC scheme can be rewritten as a three-point linear reconstruction with a specific limiter. Based on this notion, he proposed a new concept, namely steepness-adjustable limiters (with the steepness measured by a steepness parameter $\beta$). Such limiters result in the final scheme exhibiting different desired behaviors including a high-order property with exact steepness parameter values that are theoretically determined or an anti-diffusive/compression property with a larger steepness parameter. Therefore, the specific limiter in the THINC scheme belongs to this class of limiters. They further proposed a steepness-adjustable harmonic (SAH) limiter.

Based on these works, He further developed a formulation to automatically obtain the solution-dependent steepness parameter and provided a general process to design so-called self-adjusting steepness (SAS)-based schemes. The resulting schemes are effective in simulating both the smooth and discontinuous solutions with significantly improved solution quality. Such schemes are also able to effectively preserve the steepness of discontinuous solutions even after a long computation time. In such schemes, the supremum of the steepness parameter $\beta$ is not predetermined and is instead empirically determined. The objective of this study is to propose a universal method to theoretically determine the supremum of the steepness parameter in SAS schemes.

The remainder of this paper is organized as follows. In Section 1, we provide a brief review of steepness-adjustable limiters for the linear advection equation. In Section 2, we then present our analysis for obtaining the supremum of the steepness parameter $\beta$ in the steepness-adjustable limiters and provide the analytical expressions of the supremum of $\beta$ for THINC and SAH limiters. In Section 3, we propose supremum-determined self-adjusting steepness-based (SAS) schemes. In Section 4, we extend the final schemes to solve the Euler equations. The benchmarking tests for the one-dimensional linear advection equation and one- and two-dimensional Euler equations are presented in Section 5. Finally, we conclude the paper in Section 6.

# 1. Steepness-adjustable limiters for linear advection equation

The linear advection equation is the simplest example of the hyperbolic conservation law. This equation describes the transport phenomenon at a constant velocity. Considering the one dimensional case, we set $x \in \mathbb{R}$, $t \in \mathbb{R}^+$, and a constant velocity $a > 0$. The linear advection equation is:

$$\begin{cases} \partial_t u + a\, \partial_x u = 0, \\ u(x,0) = u^0(x). \end{cases} \quad (1)$$

System (1) is a closed Cauchy problem and has the following analytical solution:

$$u(x,t) = u^0(x - at).$$

## 1.1 Steepness-adjustable limiter

To perform numerical calculations, we need to discretize equation (1). Here, we adopt the following finite-difference-like discretization.

**Definition 1**. *A standard finite-difference-like discretization is:*

$$u_j^{n+1} = u_j^n - a \frac{\Delta t}{\Delta x}\left( u_{j+\frac{1}{2}}^n - u_{j-\frac{1}{2}}^n \right), \quad (2)$$

where $\Delta t$ is the time step and $\Delta x$ is the cell size. The numerical solution in cell $j$ at time step $n$ is denoted by $u_j^n$. The updated numerical solution at the next time step in the same cell is $u_j^{n+1}$.

For the linear advection equation with constant velocity (Eq.1), the numerical flux is the velocity $a$ times the half-point reconstructed value $u_{j\pm\frac{1}{2}}$. Hence, it is sufficient to only consider the reconstructed value $u_{j\pm\frac{1}{2}}$. Here, this value is constructed using the self-adjusting steepness (SAS)-based schemes with two different steepness-adjustable limiters (i.e., the THINC and SAH limiters).

**Definition 2** (SAS (THINC) scheme). *The half-point value $u_{j+\frac{1}{2}}$ reconstructed by the SAS scheme using the THINC limiter [1], when $u_{j+1} > u_j > u_{j-1}$ is*

$$u_{j+\frac{1}{2}} = u_j + \frac{1}{2}\left(1 - r_j + (1 + r_j)\frac{\tanh(\beta) + A}{1 + A\tanh(\beta)}\right)(u_{j+1} - u_j),$$

with $A = \frac{B/\cosh(\beta) - 1}{\tanh(\beta)}$, $B = \exp(\beta \frac{r_j - 1}{r_j + 1})$, $r_j = \frac{u_j - u_{j-1}}{u_{j+1} - u_j}$.

**Definition 3** (SAS (SAH) scheme). *The half-point value $u_{j+\frac{1}{2}}$ reconstructed by the SAS scheme using the SAH limiter [1] is*

$$u_{j+\frac{1}{2}} = u_j + \frac{1}{2}\frac{2r_j}{1/\beta + r_j}(u_{j+1} - u_j).$$

A typical characteristic of the steepness-adjustable limiters is that they have a steepness parameter $\beta$. This parameter enables the final scheme to exhibit different behaviors with different parameter values.

Here, we provide several important properties of these two limiters. For brevity we rewrite the THINC limiter as

$$\phi^T(r) = 1 - r + (1 + r)\frac{\tanh(\beta) + A}{1 + A\tanh(\beta)},$$

where $r$ is the ratio between the upwind increment and local increment, and the SAH limiter can be written as

$$\phi^H(r) = \frac{2r}{1/\beta + r}.$$

1) Both limiters are increasing, i.e., $\phi' \geq 0$.

2) Both limiters have the same limit if $r \to \infty$, i.e.,

$$\lim_{r \to \infty} \phi(r) = 2.$$

3) Both limiters are null at $r = 0$, i.e., $\phi(0) = 0$.

4) Both limiters are functions of $r$, i.e., $\phi'' \leq 0$

For such steepness-adjustable limiters, the steepness parameter $\beta$ is fully adjustable. The capacity for discontinuity preservation is also determined by changing the value of the steepness parameter $\beta$. The larger the value of $\beta$ the higher the capacity of the limiters for discontinuity capturing. Nevertheless, a large $\beta$ results in significant numerical oscillations which may fail the numerical calculations. In the current research works the determination of $\beta$ is empirical and a posteriori and there exists no theory to determine the supremum of the steepness parameter $\beta$.

In this study, we present a method to determine the supremum of $\beta$ for all the steepness-adjustable limiters.

## 2. Supremum analysis for steepness parameter

### 2.1 Sweby's TVD condition

Here, we provide a brief review of the limited downwind scheme proposed in [19]. The LDS scheme must satisfy the following constraints:

**Definition 4.** $u_{j+\frac{1}{2}}$ *is the closest value to the downwind value* $u_{j+1}$*, subject to the upwind stability constraints [19]:*

$$\begin{cases} m_{j+1} \leq u_{j+\frac{1}{2}} \leq M_{j+1}, \forall j \in \mathbb{Z}, \\ M_j + \frac{\Delta x}{a\Delta t}(u_j - M_j) \leq u_{j+\frac{1}{2}} \leq m_j + \frac{\Delta x}{a\Delta t}(u_j - m_j), \forall j \in \mathbb{Z}, \\ |u_{j+1} - u_{j+\frac{1}{2}}| \text{ is minimum}, \forall j \in \mathbb{Z}, \end{cases}$$

*where* $m_j = \min(u_j, u_{j-1})$, $M_j = \max(u_j, u_{j-1})$.

The abovementioned upwind stability constraints are Sweby's TVD condition [2]. As the Sweby's TVD condition involves a three-point stencil, here we reformulate it into a MUSCL-type expression with the ultrabee limiter [19]. All the limited downwind schemes should be controlled by this limiter.

**Proposition 1** (Ultrabee limiter). *For the linear advection equation (Eq.1) with constant velocity a, the ultrabee slope limiter is under the function*

$$\varphi(r_j) = \min\left(0, \max\left(2, 2\left(\frac{\Delta x}{a\Delta t} - 1\right) r_j\right)\right), \tag{3}$$

where $r_j = \frac{u_j - u_{j-1}}{u_{j+1} - u_j}$.

*Proof.* A linear reconstruction of the half-point value with the slope limiter is in the form

$$u_{j+\frac{1}{2}} = u_j + \frac{1}{2}\varphi(r_j)(u_{j+1} - u_j).$$

Thus, the slope limiter can be expressed as

$$\varphi(r_j) = 2\frac{u_{j+\frac{1}{2}} - u_j}{u_{j+1} - u_j}.$$

Again using the definition of Sweby's TVD condition, it satisfies

$$\begin{cases} m_{j+1} \leq u_{j+\frac{1}{2}} \leq M_{j+1}, \forall j \in \mathbb{Z}, \\ M_j + \frac{\Delta x}{a\Delta t}(u_j - M_j) \leq u_{j+\frac{1}{2}} \leq m_j + \frac{\Delta x}{a\Delta t}(u_j - m_j), \forall j \in \mathbb{Z}, \\ |u_{j+1} - u_{j+\frac{1}{2}}| \text{ is minimum}, \forall j \in \mathbb{Z}. \end{cases}$$

We will focus on the first and second constraints. Consider a simple stencil of three points satisfying $u_{j-1} < u_j < u_{j+1}$. Then, the inequalities can be simplified with $m_j = u_{j-1}, M_j = u_j$. The first inequality is then

$$u_j \leq u_{j+\frac{1}{2}} \leq u_{j+1},$$

which gives

$$0 \leq 2\frac{u_{j+\frac{1}{2}} - u_j}{u_{j+1} - u_j} \leq 2.$$

In the same way, the second inequality gives

$$u_j \leq u_{j+\frac{1}{2}} \leq u_{j-1} + \frac{\Delta x}{a\Delta t}(u_j - u_{j-1}),$$

which gives

$$0 \leq 2\frac{u_{j+\frac{1}{2}} - u_j}{u_{j+1} - u_j} \leq 2(\frac{\Delta x}{a\Delta t} - 1)r_j.$$

Finally, we can obtain the ultrabee limiter region

$$0 \leq \varphi(r_j) \leq \min\left(2, 2\left(\frac{\Delta x}{a\Delta t} - 1\right)r_j\right).$$

As a result, we obtain

$$\varphi(r_j) = \max\left(0, \min\left(2, 2\left(\frac{\Delta x}{a\Delta t} - 1\right)r_j\right)\right).$$

Again, we provide some fundamental properties of the ultrabee limiter.

1) Let $\frac{\Delta x}{a\Delta t} = v$, then there exists $r_0 = \frac{2v}{1-v}$ such that

$$\forall r > r_0, \varphi(r) = 2.$$

2) $\varphi(r)$ is increasing with $\varphi'' = 0$, $\forall 0 \leq r \leq r_0$.

3) $\varphi(0) = 0$.

## 2.2 Supremum analysis for steepness-adjustable limiters

Noting the Sweby's TVD slope and the steepness-adjustable limiters in Fig.1, it is seen that the steepness-adjustable limiters are downwind for sufficiently large values of $\beta$. However, to gain stability by using the TVD region [2] as closely as possible, it is necessary to adjust $\beta$ such that the steepness-adjustable limiter is completely controlled by the ultrabee limiter.

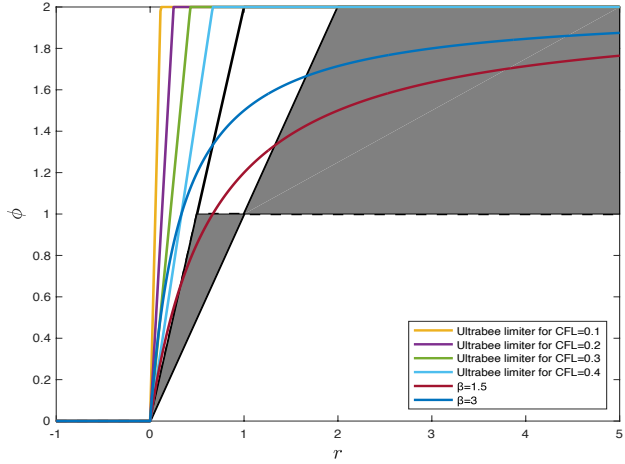

Figure 1: An example of ultrabee slopes and SAH limiters with $\beta$=1.5 and $\beta$=3

Combining the previous conclusions, it is therefore sufficient to ensure that the steepness-adjustable limiters satisfy the following two lemmas.

**Lemma 1**.

$$\forall r > r_0, \varphi(r) \geq \phi(r)$$

*Proof.*

$$\forall r > r_0, \lim_{r \to \infty} \phi(r) = 2 = \varphi(r).$$

Lemma 1 can be easily derived from the sign-preserving property of the limit.

**Lemma 2**.

$$\forall 0 \leq r \leq r_0 \text{ if } \varphi'(0) \geq \phi'(0), \text{ then } \varphi(r) \geq \phi(r).$$

*Proof.* Now, consider the function defined by $\psi \equiv \varphi - \phi$. From the hypothesis of Lemma 2, $\psi'(0) \geq 0$. It is obvious that $\psi(0) = 0$. $\forall 0 \leq r \leq r_0$,

$$\psi(r) = \phi(0) + \int_0^r \psi'(s)ds$$
$$= \psi(0) + \int_0^r \left[\psi'(0) + \int_0^s \psi''(l)dl\right]ds$$
$$= \psi(0) + \psi'(0)r + \int_0^r \left[\int_0^s \psi''(l)dl\right]ds$$

$\varphi''(r) = 0$; thus, $\psi''(r) = \varphi''(r) - \phi''(r) = -\phi''(r)$. In addition, both steepness-based limiters are concave, that is, $\phi''(r) \leq 0$, which gives $\psi''(r) \geq 0$. Therefore,

$$\psi(r) = \underbrace{\psi(0)}_{=0} + \underbrace{\psi'(0)r}_{\geq 0} + \int_0^r \left[\underbrace{\int_0^s \psi''(l)\,dl}_{\geq 0}\right]ds \geq 0,$$

which gives $\varphi(r) \geq \phi(r), \forall 0 \leq r \leq r_0$.

## 2.3 The supremum-determined THINC limiter and SAH limiter

In this section, we apply Lemmas 1 and 2 to the SAH and THINC limiters and obtain the supremum of the steepness parameter.

**Theorem 1**. *The supremum of $\beta$ in SAH limiter for the linear advection equation is*

$$\beta \leq \frac{1}{\nu} - 1.$$

*Proof.* This proof follows directly from Lemma 2 as the first-order derivation of the SAH limiter is simply calculated.

**Theorem 2**. *The supremum of $\beta$ in the THINC limiter for the linear advection equation is*

$$\beta \leq \frac{1}{2}\left(\frac{1}{\nu} - 1\right).$$

*Proof.*

$$(\phi^T)'(0) = -2 + \frac{2\beta e^{-\beta}}{\sinh(\beta)} \frac{1 - \tanh^2(\beta)}{(1 - \tanh(\beta))^2}$$

$$= -2 + \frac{2\beta}{\sinh(\beta)e^{-\beta}} = -2 + \frac{4\beta}{1 - e^{-2\beta}}.$$

Now, applying lemma 2,

$$-2 + \frac{4\beta}{1 - e^{-2\beta}} \leq 2\left(\frac{1}{\nu} - 1\right)$$

$$\frac{2\beta}{1 - e^{-2\beta}} \leq \frac{1}{\nu}$$

$$2\beta + \frac{1}{\nu}\frac{1}{1 + \sum_{i=1}^{\infty}\frac{(2\beta)^i}{i!}} \leq \frac{1}{\nu}.$$

To simplify the final expression of $\beta$, a sufficient condition is used:

$$2\beta + \frac{1}{\nu}\frac{1}{1 + \sum_{i=1}^{\infty}\frac{(2\beta)^i}{i!}} \leq 2\beta + \frac{1}{\nu}\frac{1}{1 + 2\beta} \leq \frac{1}{\nu}.$$

which gives

$$\beta \leq \frac{1}{2}\left(\frac{1}{\nu} - 1\right).$$

The obtained supremum of $\beta$ is however underestimated. This is because the high-order terms in the exponential series are ignored. To consider the high-order terms, we introduce a correction function $\eta^c(\nu)$ which is defined as the ratio between the exact numerical solution and the theoretical solution proposed here, $\beta^t = \frac{1}{2}\left(\frac{1}{\nu} - 1\right)$. The exact numerical solution is obtained by solving the transcendental equation

$$2\nu\beta + e^{-2\beta} - 1 = 0$$

using the Newton dichotomy algorithm. The results are shown in Figs. 2-4.

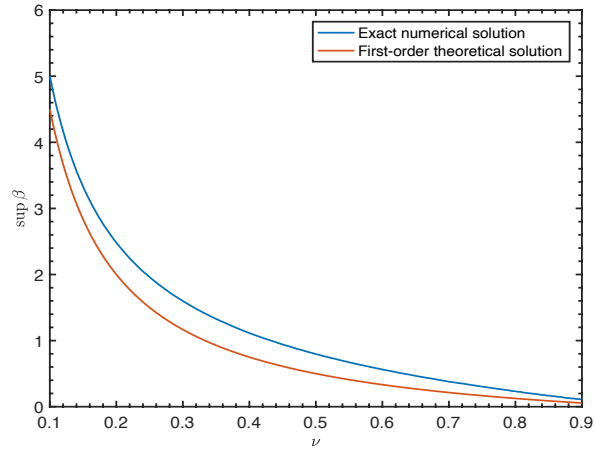

Figure 2: Exact numerical solution $\beta^n$ and corrected theoretical numerical solution $\beta^c$

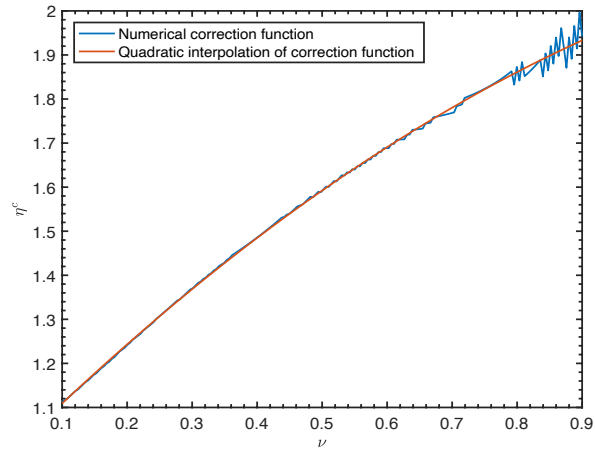

Figure 3: Exact numerical solution $\beta^n$ and corrected theoretical numerical solution $\beta^c$

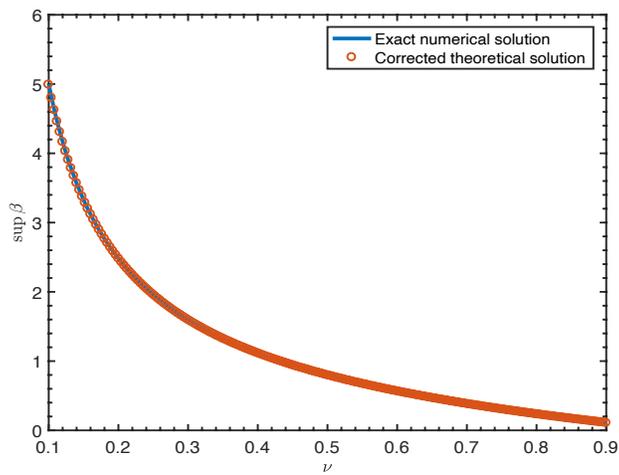

Figure 4 : Exact numerical solution $\beta^n$ and corrected theoretical numerical solution $\beta^c$

Here, we provide the quadratic fitting of the exact numerical correction function

$$\eta^c(v) = -0.4441v^2 + 1.474v + 0.9662.$$

Using this we can finally obtain the corrected Theorem 2.

**Proposition 2** (Corrected Theorem 2). *The supremum of β in the THINC limiter for the linear advection equation is*

$$\beta \leq \frac{\eta^c(v)}{2}\left(\frac{1}{v} - 1\right)$$

*with*

$$\eta^c(v) = -0.4441v^2 + 1.474v + 0.9662.$$

As it is seen the supremum of $\beta$ depends on the Courant-Friedrichs-Lewy number $v$. Therefore, the steepness can be directly determined by the CFL number. In Table 1, we present the corresponding CFL numbers according to several typical values of $\beta$ in the THINC limiter. We also provide the corresponding steepness parameter in the SAH limiter. These results confirm the observation made in [1] that the THINC (with $\beta \simeq 2.1$) limiter performs the same as the SAH (with $\beta \simeq 2.9$) limiter.

| β THINC    | 0.23 | 0.56 | 1.11 | 1.6  | 2.1  | 2.3  |
|------------|------|------|------|------|------|------|
| CFL number | 0.80 | 0.60 | 0.40 | 0.30 | 0.26 | 0.21 |
| β SAH      | 0.25 | 0.67 | 1.50 | 2.3  | 2.9  | 3.8  |

Table 1 : Correspondence between β of THINC, β of SAH, and CFL number

## 3. Supremum-determined self-adjusting steepness based schemes

In ref.[1], He proposed an analytical approach to automatically determine the solution-dependent steepness parameter and further provided a general process to design the self-adjusting steepness (SAS) based schemes. The resulting schemes are capable of simulating both smooth and discontinuous solutions with significantly improved solution quality. Such schemes can also effectively preserve the steepness of discontinuous solutions even after a long computation time. According to ref.[1], the final steepness parameter is

$$\beta = \xi_j \beta_s + (1 - \xi_j)\beta_l$$

where $\xi_j = \frac{4r_j^4}{(r_j^4+1)^2}$ and only $\beta_s$ and $\beta_l$ are required to be determined. Here, $\beta_s$ represents the value at which the THINC/SAH limiter acts as a second-order limiter. For the SAH scheme, $\beta_s = 1$ is transformed to the classical van Leer limiter. The THINC limiter is also a second-order limiter for $\beta_s = \ln(3)$, as shown in [1].

However, $\beta_l$ is a posteriori in [1]. In contrast to the previous works, here we obtain $\beta_l$ a priori by taking its supremum value. Thus, we set $\beta_l = \frac{\eta^c(v)}{2}(\frac{1}{v} - 1)$ for the THINC limiter, and $\beta_l = \frac{1}{v} - 1$ for the SAH limiter. The final SAS schemes are called supremum-determined SAS schemes.

The SAH limiter with the steepness parameter equal to the supremum as in Theorem 1 is referred to as the SAH (supremum). Similarly, the THINC limiter with the steepness parameter equal to the supremum as in the corrected Theorem 2 is referred to as the THINC (supremum). In addition, we denote the final SAS scheme using SAH and THINC limiters as SAS (SAH) and SAS (THINC), respectively.

## 4. Extension to solve the Euler equations

In this section, we extend the final schemes to solve the compressible Euler equations. Such solutions are essential in scientific studies and engineering applications.

### 4.1 On the spatial discretization

The one-dimensional compressible Euler equations are:

$$\mathbf{U}_t + \mathbf{F}(\mathbf{U})_x = 0, \tag{4}$$

where

$$\mathbf{U} = [\rho, \rho u, E]^T,$$
$$\mathbf{F}(\mathbf{U}) = [\rho u, \rho u^2 + p, (E+p)u]^T.$$

This set of equations describes the conservation law of mass density $\rho$, momentum density $\rho u$, and total energy density $E = \rho e + \frac{1}{2}\rho u^2$, where $e$ denotes the internal energy per unit mass. To close this set of equations, the ideal gas equation of state $p = (\gamma - 1)\rho e$ is used, where $\gamma$ is the ratio of the specific heats (5/3 for the perfect gas and 1.4 for the air).

The one-dimensional Euler equations comprise three different characteristic fields. The characteristic speeds are different over the computational domain. However, the new supremum-determined THINC/SAH scheme is extremely dependent on the local characteristic speeds. This is because the steepness parameter is based on $\nu$. Therefore, to enable the proposed scheme to be applicable in real applications, we must use a reasonable value for $\nu$ such that the steepness parameter takes the same value over the computational domain and in each time step.

For the one-dimensional Euler equations, the characteristic speeds in cell $i$ are smaller than $|u_i| + |c_i|$, where $c_i$ is the local sound speed. Thus, all the characteristic speeds over the computational domain must be smaller than $a_{\max} = \max_i |u_i| + |c_i|$. Using this value and the previous fixed CFL number $\nu_0$, the time step is then determined through the CFL condition.

**Definition 5** (CFL condition). *For the one-dimensional Euler equations, the CFL condition provides the time step which is*

$$\Delta t = \nu_0 \frac{\Delta x}{a_{\max}}.$$

With the time step obtained by the CFL condition, we also notice that the local CFL number over the computational domain must be smaller than $\nu_0$ $\nu = \frac{a \Delta t}{\Delta x} = \frac{a}{a_{\max}} \nu_0 \leq \nu_0$. Since $\nu \leq \nu_0$ (from Fig.1) the THINC/SAH limiter controlled by $\nu_0$-ultrabee slope must be controlled by $\nu$-ultrabee slope. Therefore, we only need to calculate the steepness parameter with $v_0$ to ensure that the whole computational domain satisfies Theorem 1 (or Theorem 2).

To apply the conclusions, we made for the advection linear equation to the Euler equations, we further need to decouple the Euler equations to its characteristic space. This enables the decoupled system to be approximated as three advection equations.

The flux-split-based FDM often comprises of the following two steps. The first step splits the flux vector into positive and negative flux vectors, $\mathbf{F}_j^+$ and $\mathbf{F}_j^-$, as $\mathbf{F}_j = \mathbf{F}_j^+ + \mathbf{F}_j^-$, where the two Jacobians $\frac{\partial \mathbf{F}^+(\mathbf{U})}{\partial \mathbf{U}}|_j$ and $\frac{\partial \mathbf{F}^-(\mathbf{U})}{\partial \mathbf{U}}|_j$ are diagonalizable with non-negative/non-positive eigenvalues, respectively [15,24–26]. The second step is to use the numerical schemes to obtain the numerical flux $\widetilde{\mathbf{F}}_{j\pm1/2}$, where we employ a local characteristic decomposition technique as the following.

Firstly, the positive and negative flux vectors are projected onto the characteristic space: $\mathbf{W}_{j+l}^\pm = \mathbf{L}(\mathbf{U}_{i+1/2})\mathbf{F}_{i+l}^\pm$, where $\mathbf{U}_{i+1/2}$ is the Roe average of $\mathbf{U}_i$ and $\mathbf{U}_{i+1}$, $l = -q, -q+1, \ldots, q$, and $q$ is a positive number depending on the numerical stencil. Furthermore, $\mathbf{L}$ is the matrix formed by the left eigenvectors according to the state $\mathbf{U}_{i+1/2}$. Here we use the numerical schemes to obtain the reconstructed characteristic numerical flux $\widetilde{\mathbf{W}}_{j\pm1/2}$.

The last step is to reproject the constructed characteristic numerical flux back to the physical space to obtain the final numerical flux,

$$\widetilde{\mathbf{F}}_{j\pm1/2} = \mathbf{R}(\mathbf{U}_{i+1/2})\widetilde{\mathbf{W}}_{j\pm1/2},$$

where $\mathbf{R}$ is the matrix formed by the right eigenvectors according to the state $\mathbf{U}_{i+1/2}$. For instance, the final semi-discretizing equation in the x-direction is

$$\frac{d\mathbf{U}_j}{dt} = -\frac{1}{\Delta x}(\widetilde{\mathbf{F}}_{j+1/2} - \widetilde{\mathbf{F}}_{j-1/2}).$$

The characteristic fields can be also divided into genuinely nonlinear, and linearly degenerate fields. The SAS scheme might apply to all fields. However, the phenomenon of transforming the square Gauss to the step function has not been completely avoided in the SAS scheme. This is because we use $\xi$ which is determined through the WENO methodology. As stated in [27], high gradients and fine smooth structures may be regarded as discontinuities by the WENO smoothness indicators in cases where the grids are relatively coarse. Therefore, we follow Harten's [28] suggestion that the general TVD scheme has good performance in genuinely nonlinear characteristic fields, and the SAS scheme is applied only to the linearly degenerate fields. In this paper, by using the the TVD scheme in the nonlinear characteristic fields we present the numerical results with the CFL number larger than 0.5 for the Euler equations. For the linear advection and the nonlinear burgers equations, we only present the numerical results for the CFL numbers smaller than 0.5 as suggested in the related literature [19,20].

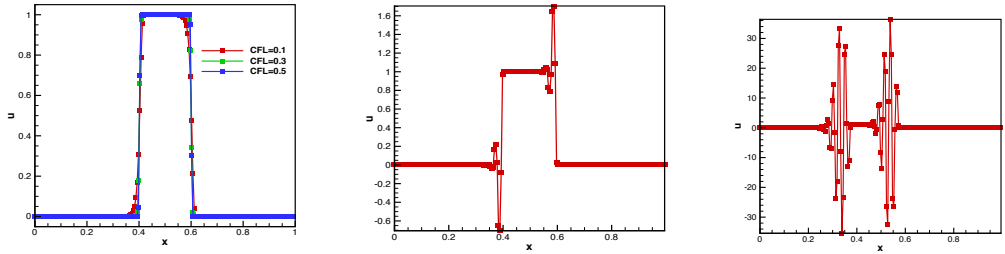

Figure 5: Numerical results of step test in section 5.1.1 on a 201-point grid at *t=5* computed by MUSCL scheme. (left) *v=0.1,0.3,0.5*; (middle) *v=0.55*; (right) *v=0.6*.

*4.2 On the time integration*

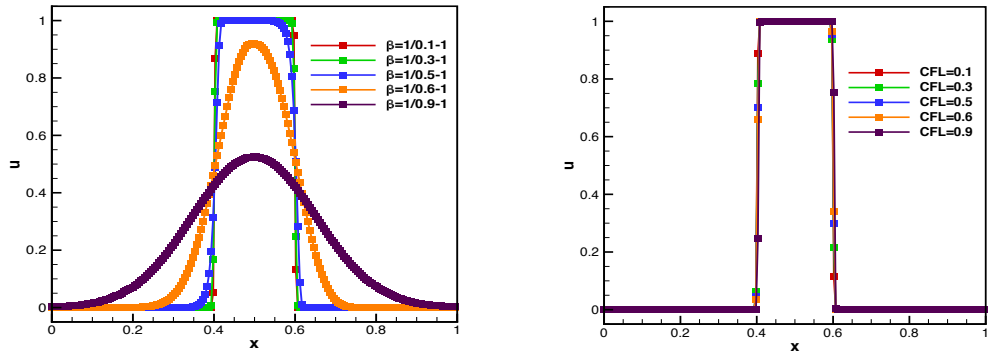

Figure 6: Numerical results of step test in section 5.1.1 on a 201-point grid at *t=5* computed by (left) SAH scheme with a fixed time step $\Delta t = 1 \times 10^{-4}$ and (right) SAH (supremum) scheme with the time step calculated by CFL condition.

It can be easily seen that the new spatial scheme is a CFL-dependent scheme. The definition of the CFL number implies that discretization in the spatial and temporal domains should be closely coupled. Therefore, it is essential to consider spatial discretization and temporal discretization simultaneously.

1) The time integration scheme greatly affects the performance of the spatial scheme and introduces negative dissipation. Fig.5 shows the numerical results of step test for advection equation computed by MUSCL scheme [11,12] under different CFL numbers. It is seen that the resolution of the discontinuities varies with different CFL numbers. The total numerical dissipation is also decreased by increasing the CFL number. For a CFL number larger than 0.5, the numerical results become unstable. This suggests that the time integration scheme directly affects the performance of the spatial scheme. As reported in literature [29], the combination of the explicit Euler scheme and high order spatial scheme is unstable. It is also seen in Fig.7 that the dissipation of the explicit Euler scheme is negative as it is decreased by increasing $\Delta t$.

2) The conclusions derived above are the strictest stability requirements for the SAH (supremum) scheme.

To exclude the effect of the temporal discretization, we further conduct several numerical tests for the advection equation using a fixed but very small time step. Fig.9 (left) shows the numerical results of step test for advection equation computed by the SAH scheme with different steepness parameters, where $\Delta t = 1 \times 10^{-4}$. In this condition, the time dissipation is fixed to be very small. The steepness parameter is also chosen using the analytical result with the CFL number set to 0.1, 0.3, 0.5, 0.6, and 0.9. The numerical results in Fig.9 (left) show that the numerical dissipation is decreased by increasing the CFL number. Fig.9 (right) shows that the numerical results of the SAH (supremum) scheme for different CFL numbers. It is seen that the numerical results are very sharp for a large CFL scale, from 0.1 to 0.9. These suggest that the combination of the new scheme and the explicit Euler scheme ensures a minimal total dissipation for various CFL numbers.

3) The stability requirements derived above can be slightly relaxed by using high-order temporal discretization methods.

From the discussions above, it is seen that for a small enough temporal dissipation, the numerical dissipation of the SAH is decreased by increasing the value of steepness parameters. A naive way to reduce temporal dissipation is to use high-order temporal discretization methods. Therefore, when the SAH/THINC (SAH) scheme is applied to the compressible Euler equations, the third-order Runge-Kutta time integration scheme is recommended to achieve higher numerical stability.

## 5. Numerical tests

### 5.1 Linear advection equation

Here, we consider a one-dimensional linear advection equation Eq.1 with $a = 1$. The numerical solutions are obtained for different CFL numbers. An explicit Euler scheme is used for time discretization.

*5.1.1 Step test*

First, we present a step-function test. The initial condition is an indicatrix function in the interval [0,1]:

$$u(x, 0) = \begin{cases} 1, & \text{if } 0.4 \leq x \leq 0.6, \\ 0, & \text{otherwise,} \end{cases}$$

with a 201-point grid, and the final time is $t = 5$. A periodic boundary condition is also applied.

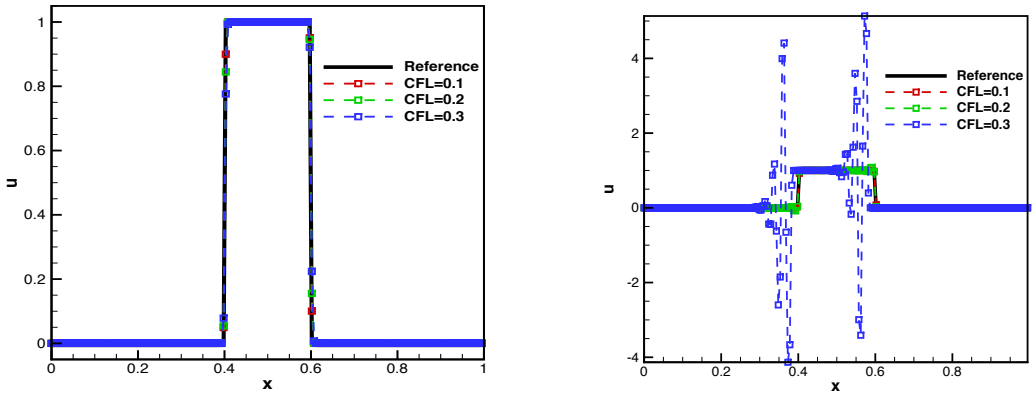

Figure 7: Numerical results of THINC (supremum) scheme on a 201-point grid at $t=5$. (left) β is the supremum given in corrected Theorem 2; (right) β is 1.5 times the supremum given in corrected Theorem 2.

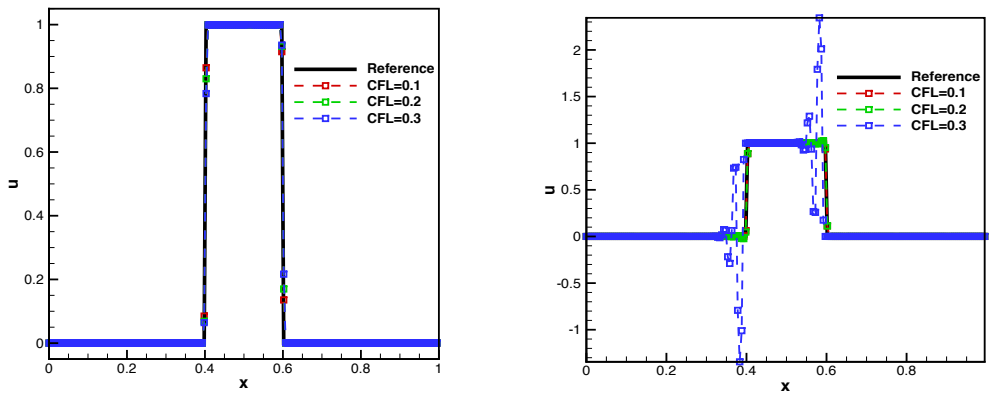

Figure 8: Numerical results of SAH (supremum) scheme on a 201-point grid at $t=5$. (left) β is the supremum given in Theorem 1; (right) β is 1.5 times the supremum given in Theorem 1.

Fig.7 depicts the numerical solutions obtained by the THINC (supremum) scheme for cases where the supremum of $\beta$ is and is not satisfied and for different CFL numbers. Fig.8 depicts the numerical solutions obtained by the SAH (supremum) scheme where the supremum of $\beta$ is and is not satisfied for different CFL numbers.

As it is seen for all CFL numbers, where the supremum of $\beta$ is theoretically determined, the THINC (supremum) and SAH (supremum) schemes have no numerical oscillations. The numerical solutions are also in good agreement with the exact solution. However, when the supremum of $\beta$ does not satisfy the constraints, a significant numerical oscillation is seen for both schemes which agree with our analytical results.

*5.1.2 Harten test*

The Harten test [30] includes discontinuities, discontinuities of derivatives, and smooth regions in the interval $[-1,1]$. The initial conditions are as follows:

$$u(x,0) = \begin{cases} 2x + 2 - \frac{1}{6}\sin\left(3\pi\left(x - \frac{1}{2}\right)\right), & \text{if } -1 \leq x < -\frac{1}{2}, \\ \left(\frac{1}{2} - x\right)\sin\left(\frac{3\pi}{2}\left(x - \frac{1}{2}\right)^2\right), & \text{if } -\frac{1}{2} \leq x < \frac{1}{6}, \\ |\sin(2\pi(x - \frac{1}{2}))|, & \text{if } \frac{1}{6} \leq x < \frac{5}{6}, \\ 2x - 2 - \frac{1}{6}\sin\left(3\pi\left(x - \frac{1}{2}\right)\right), & \text{otherwise.} \end{cases}$$

Here, for different CFL numbers, we present the results obtained after 2 and 20 revolutions for 101 and 501 cells, respectively, by the SAH (supremum) scheme (in Figs.9-10), THINC (supremum) scheme (in Figs.11-12), SAS (SAH) scheme (in Figs.13-14), and SAS (THINC) scheme (in Figs.15-16).

In Fig.9 and Fig.11, using the supremum of $\beta$, the final scheme converts an initially continuous solution to a step-like solution. This suggests the over-compression phenomenon as described in [19].

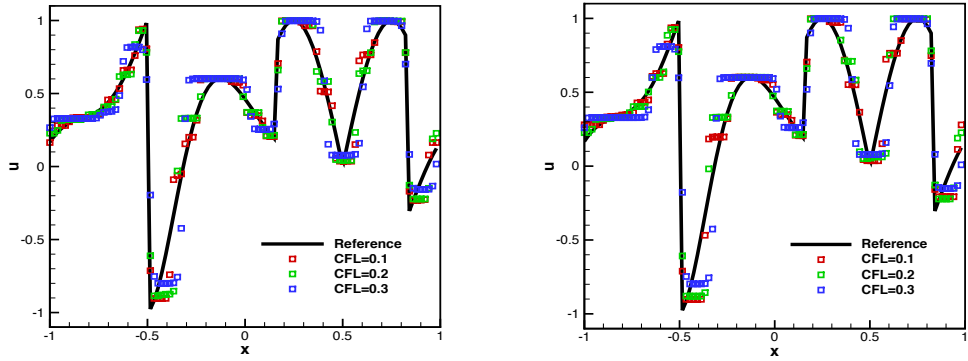

Figure 9: Numerical results of SAH (supremum) scheme on a 101-point grid. (left) at *t=2*; (right) at *t=20*

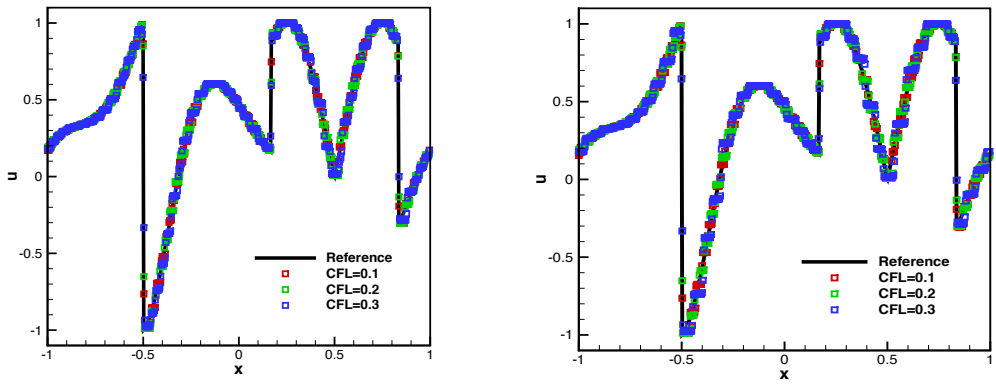

Figure 10: Numerical results of SAH (supremum) scheme on a 501-point grid. (left) at *t=2*; (right) at *t=20*.

The results for 501 cells also show that the scheme is convergent (Fig.10 and Fig.12). The width of the steps is also decreased by increasing the cell numbers.

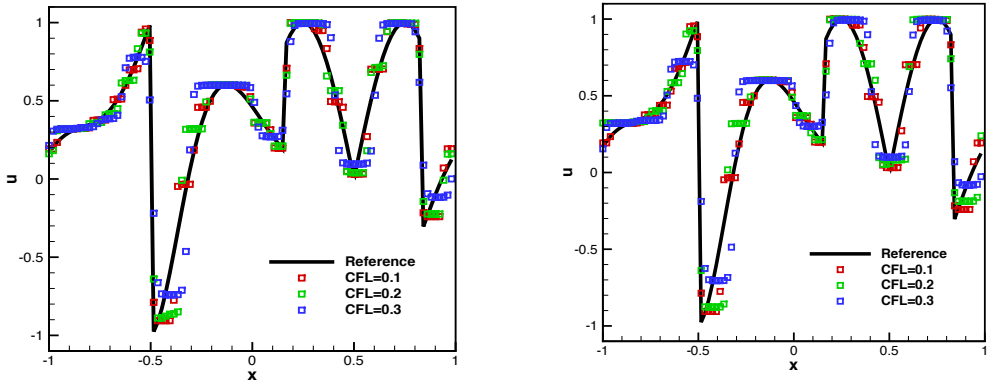

Figure 11: Numerical results of THINC (supremum) scheme on a 101-point grid. (left) at $t=2$; (right) at $t=20$.

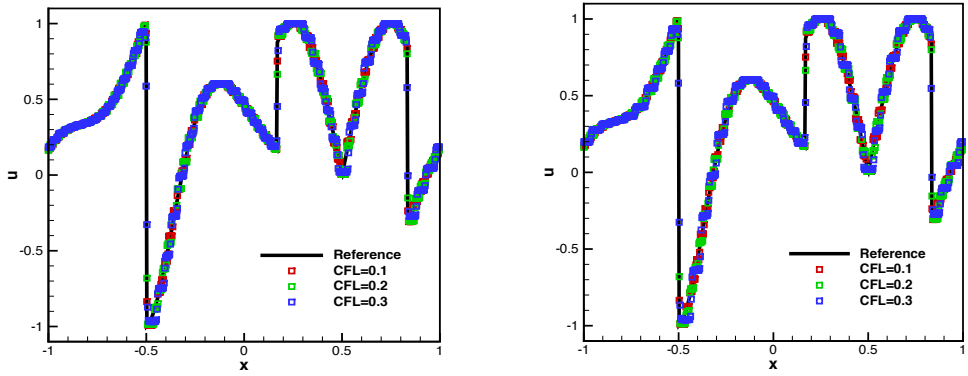

Figure 12: Numerical results of THINC (supremum) scheme on a 501-point grid. (left) at $t=2$; (right) at $t=20$.

In Figs.13-16, we present the numerical results using SAS (SAH) and SAS (THINC), where $\beta$ is set as in Theorem 1 for the SAH and in the corrected Theorem 2 for THINC. It is seen that the final results obtained by the SAS schemes exhibit the over-compression phenomenon even for small CFL numbers. This is caused by the use of the Euler explicit time scheme [29]. It is also seen in the following that if using the third-order Runge-Kutta scheme [15], the SAS schemes achieve higher performance. However, in this study, our focus is on the impact of the steepness parameter on the SAS/THINC schemes. Therefore, we left the coupling between the time discretization and the space discretization schemes for later studies.

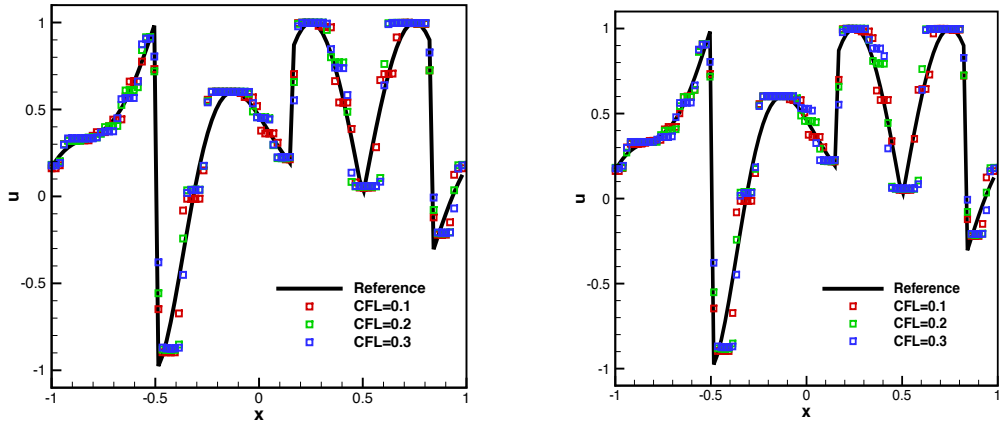

Figure 13: Numerical results of SAS (SAH) scheme on a 101-point grid. (left) at *t=2*; (right) at *t=20*.

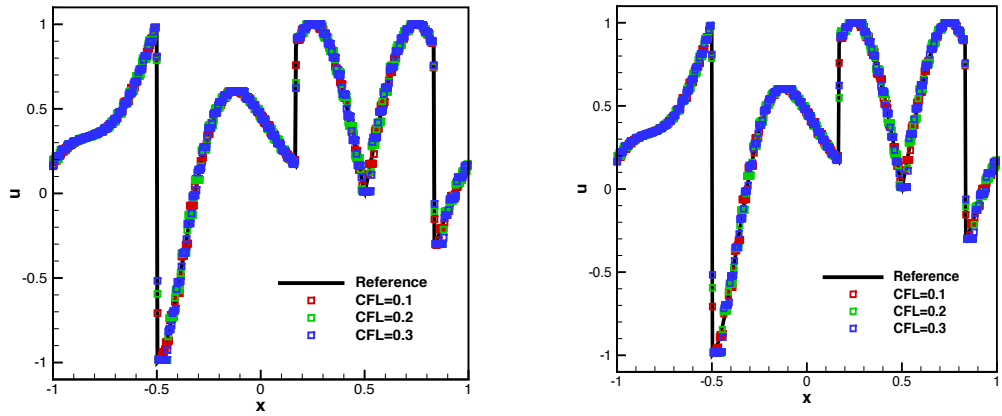

Figure 14: Numerical results of SAS (SAH) scheme on a 501-point grid. (left) at *t=2*; (right) at *t=20*

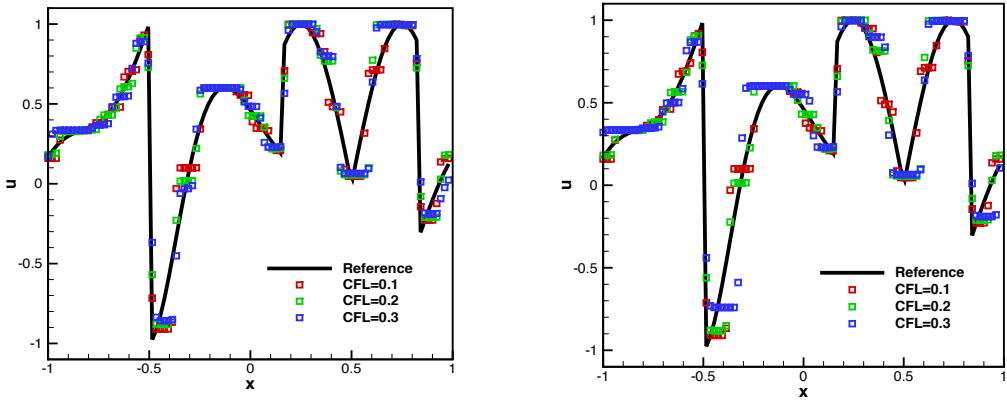

Figure 15: Numerical results of SAS (THINC) scheme on a 101-point grid. (left) at *t=2*; (right) at *t=20*

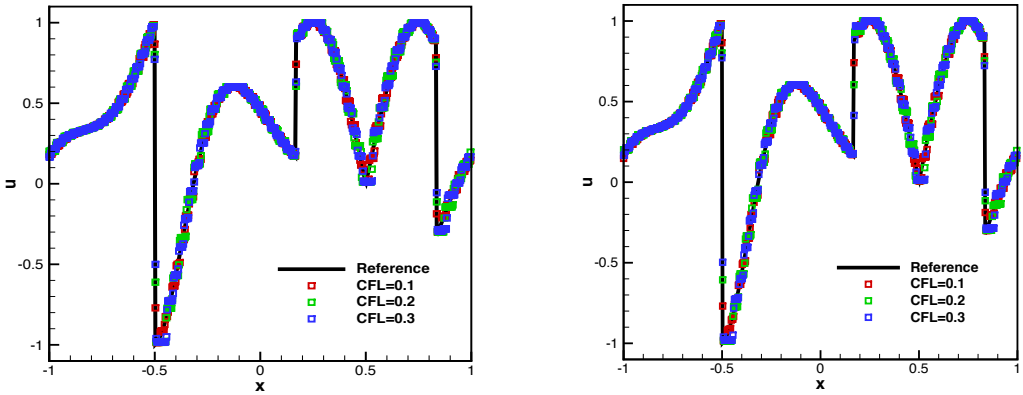

Figure 16: Numerical results of SAS (THINC) scheme on a 501-point grid. (left) at *t=2*; (right) at *t=20*

### 5.1.3 Jiang and Shu's test

Here we also perform Jiang and Shu's test [17]. The initial condition of this test includes discontinuities and a smooth profile. We set the initial condition as

$$u(x,0) = \begin{cases} \frac{1}{6}(G(x,\beta,z-\delta) + G(x,\beta,z+\delta) + 4G(x,\beta,z)), & -0.8 \leq x \leq -0.6, \\ 1, & -0.4 \leq x \leq -0.2, \\ 1 - |10(x-0.1)|, & 0 \leq x \leq 0.2, \\ \frac{1}{6}(F(x,\alpha,a-\delta) + F(x,\alpha,a+\delta) + 4F(x,\alpha,a)), & 0.4 \leq x \leq 0.6, \\ 0, & \text{otherwise}, \end{cases}$$

where the computation domain is $[-1,1]$. The functions $F$ and $G$ are defined as

$$G(x,\beta,z) = \exp(-\beta(x-z)^2), F(x,\alpha,a) = \sqrt{\max(1-\alpha^2(x-a)^2, 0)},$$

and the coefficients to determine the initial profile are given by

$$a = 0.5, z = 0.7, \delta = 0.005, \alpha = 10, \beta = \ln(2)/(36\delta^2).$$

For different CFL numbers, here we present the results obtained after 10 revolutions for 201 and 501 cells, respectively, using the SAH (supremum) scheme (in Fig.17), THINC (supremum) scheme (in Fig.18), SAS (SAH) scheme (in Fig.19), and SAS (THINC) scheme (in Fig.20).

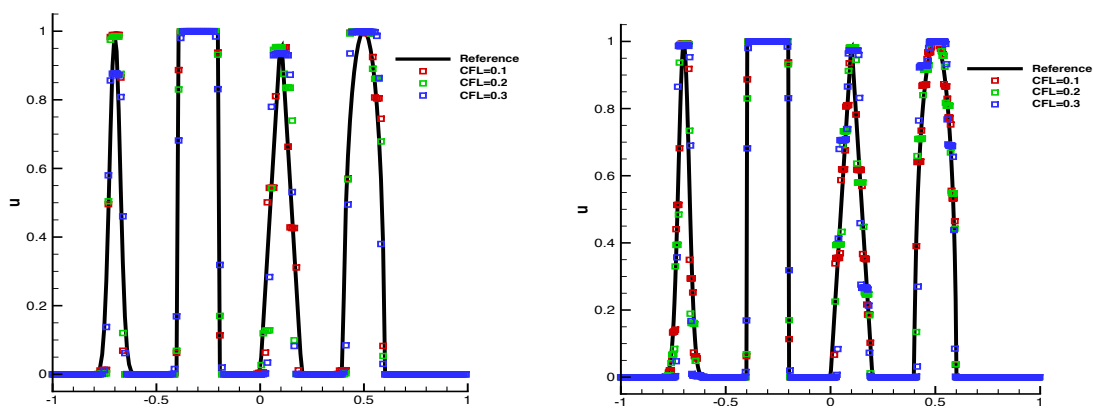

Figure 17: Numerical results of Jiang and Shu's test after ten periods (*t=20*) using SAH (supremum) scheme; (left) on 201 grid; (right) on 501 grid.

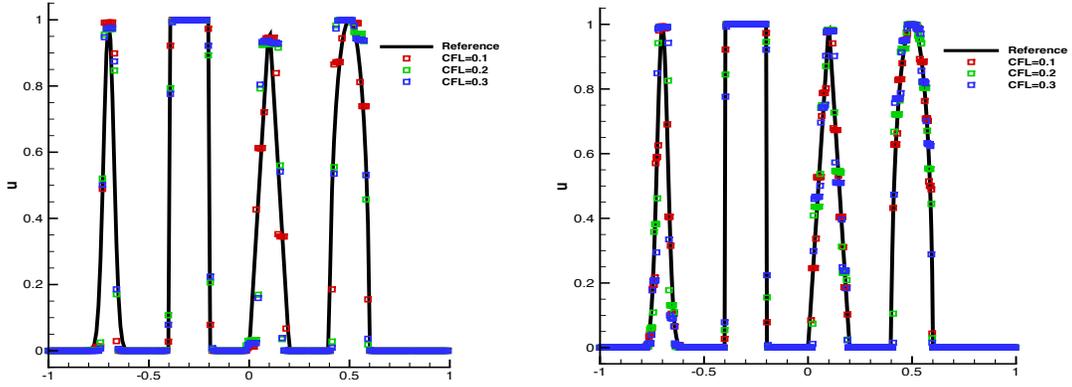

Figure 18: Numerical results of Jiang and Shu's test after ten periods (*t=20*) using THINC (supremum) scheme; (left) on 201 grid; (right) on 501 grid.

It is seen that the SAH (supremum) and THINC (supremum) schemes do not show numerical oscillations for all CFL numbers. This confirms the validity of Theorem 1. The numerical solutions on the 501-point grid are also converged.

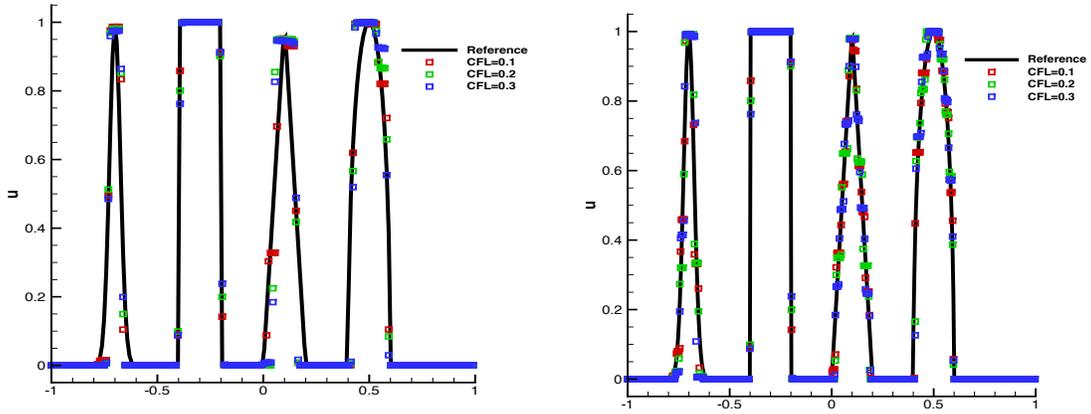

Figure 19: Numerical results of Jiang and Shu's test after ten periods (*t=20*) using SAS (SAH) scheme; (left) on 201 grid; (right) on 501 grid.

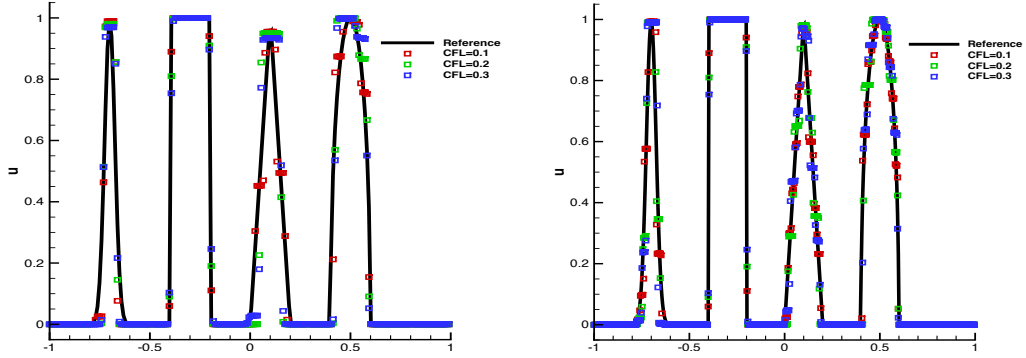

Figure 20: Numerical results of Jiang and Shu's test after ten periods (*t=20*) using SAS (THINC) scheme; (left) on 201 grid; (right) on 501 grid.

The numerical tests above suggest two ways to eliminate numerical oscillations caused by the over-compression phenomenon:

- For a fixed CFL number, the steepness parameter $\beta$ must be adjusted to satisfy Theorem 1;

- For a fixed steepness parameter $\beta$, the CFL number must be decreased until Theorem 1 is satisfied.

Note that the observed over-compression phenomenon even for smaller CFL numbers, is caused by the Euler explicit time scheme [29].

## *5.2 Burgers equation*

In this section, we solve the non-linear scalar Burgers equation

$$\frac{\partial u}{\partial t} + \frac{\partial}{\partial x}\left(\frac{u^2}{2}\right) = 0, \tag{5}$$

with the initial condition $u(x, t = 0) = 1 + 0.5\sin(\pi x)$ and 201 uniform grid points in $[-1,1]$ with a 2-periodic boundary condition and $t = 1.5$. The time discretization scheme is based on the Euler explicit scheme.

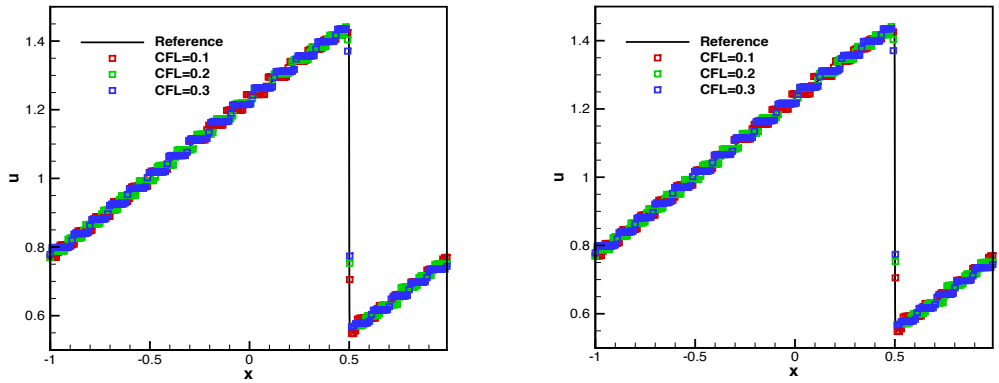

Figure 21: Numerical results of Burgers equation. (left) using SAH (supremum); (right) using THINC (supremum)

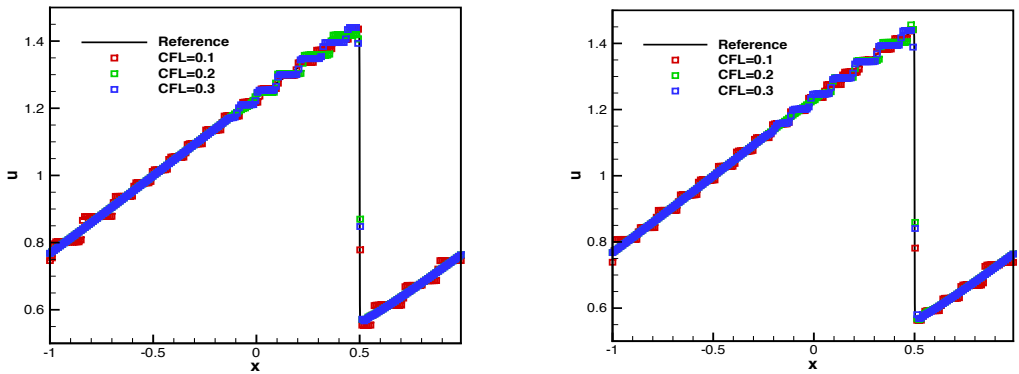

Figure 22: Numerical results of Burgers equation. (left) using SAS (SAH); (right) using SAS (THINC)

The numerical results obtained using the SAH (supremum) and THINC (supremum) schemes are plotted in Fig.21, and those of the SAS (SAH) and SAS (THINC) schemes are shown in Fig.22. It is seen that if the supremum of the steepness is respected, there are no numerical oscillations. However, as in the advection tests, the over-compression phenomenon exists which confirms the conclusions stated previously.

## 5.3 One-dimensional Euler equations

In this section, we perform five tests for one-dimensional Euler equations using four different schemes including THINC (supremum), SAH (supremum), SAS (THINC), and SAS (SAH). The third-order TVD Runge-Kutta scheme is used for time discretization. The reference solution is also obtained using the WENO5 scheme on a 4001-point grid unless otherwise stated. For every test, the supremum of $\beta$ is determined for the corresponding CFL number. For each test we obtain the results for three different CFL numbers, $\nu = \frac{1}{4}, \frac{1}{3}, \frac{2}{3}$.

### 5.3.1 Sod problem

The initial condition for the Sod shock tube is [31]

$$(\rho, u, p) = \begin{cases} (1,0,1), & \text{if } 0 \leq x \leq 0.5 \\ (0.125, 0, 0.1), & \text{if } 0.5 \leq x \leq 1, \end{cases}$$

and $t = 0.2$. A zero-gradient boundary condition is also applied at $x = 0$ and $x = 1$. Figs.23-24 present the computed density distributions obtained using different numerical schemes for different CFL numbers. As expected all the obtained solutions are in good agreement with the reference solution for the right shock and left rarefaction wave since they are all computed by the TVD scheme.

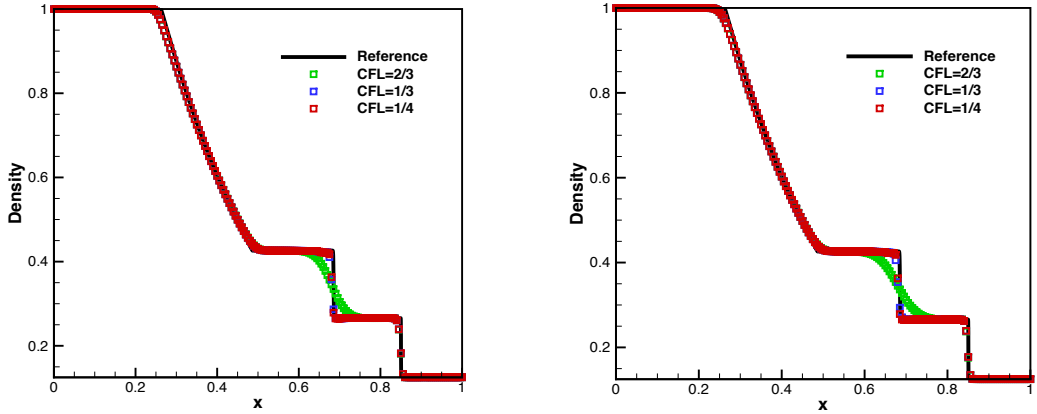

Figure 23: Sod problem with RK3. (left) using SAH (supremum); (right) using THINC (supremum).

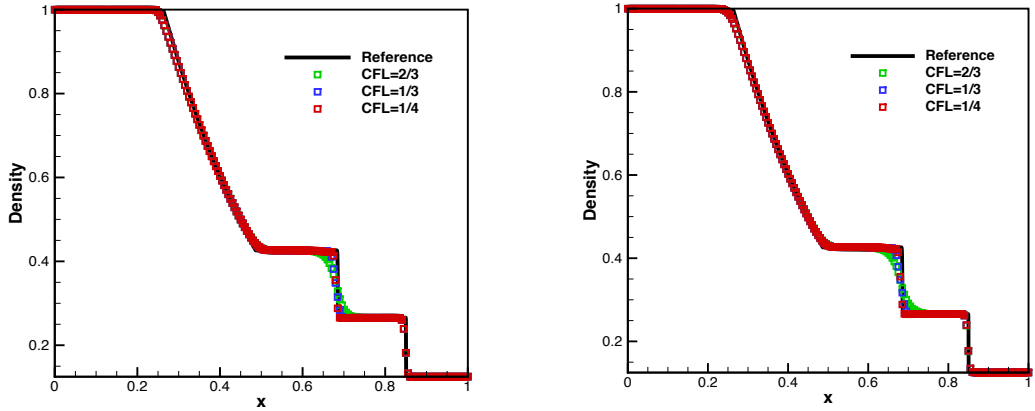

Figure 24: Sod problem with RK3. (left) using SAS (SAH); (right) using SAS (THINC).

The main difference is in the contact discontinuity, where steepness-adjustable schemes are applied.

1) For all numerical schemes, if the CFL number is larger than 0.5, there is a great dissipation at the contact discontinuity. A possible reason is that the steepness-adjustable limiters lay mainly in the TVD diffusive region when $v > 0.5$. As with all TVD schemes, the THINC (supremum) and SAH (supremum) schemes cannot effectively capture the contact discontinuity at this stage.

2) However, for $\nu < 0.5$, the THINC (supremum) and SAH (supremum) schemes show a good capacity for discontinuity preservation, as shown in Fig.23. The results obtained by these schemes are also in good agreement with the reference solutions. The capacity for discontinuity preservation increases with a decrease in the CFL number which is also in line with our theoretical results. In addition, when the self-adjusting approach of $\beta$ is employed, the numerical dissipation is slightly increased, but it does not influence the sharp contact discontinuity for small CFL numbers, such as $\nu = 0.25$.

| Method | MUSCL | SAH (supremum) | THINC (supremum) |
|---|---|---|---|
| CPU time (ms) | 4108.092 | 4349.050 | 6276.394 |
| Method | | SAS (SAH) | SAS (THINC) |
| CPU time (ms) | | 8756.082 | 12454.497 |

Table 2: Comparison of CPU times of different numerical schemes for the Sod test on 5001 grid points with $\nu=1/4$.

Table 2 compares the CPU times needed for the numerical simulation of the Sod test under $\nu = 1/4$ by using the MUSCL, SAH (supremum), THINC (supremum), SAS (SAH), and SAS (THINC) schemes. As it is seen, the CPU time required for the SAH (supremum) scheme is very close to that for the MUSCL scheme. However, the CPU time for the THINC (supremum) scheme is longer than that for the SAH (supremum) and MUSCL schemes. If the SAS schemes are employed, the CPU time is almost doubled. However, compared with the time cost of the MUSCL scheme, this time cost is reasonable.

*5.3.2 Lax problem*

The second considered problem is the Lax problem [14]. The initial condition is

$$(\rho, u, p) = \begin{cases} (0.445, 0.698, 3.528), & \text{if } 0 \leq x \leq 0.5 \\ (0.5, 0, 0.571), & \text{if } 0.5 \leq x \leq 1. \end{cases}$$

and $t = 0.16$. A zero-gradient boundary condition is applied at $x = 0$ and $x = 1$. Figs.25 -26 depict the computed density distributions using different numerical schemes under different CFL numbers.

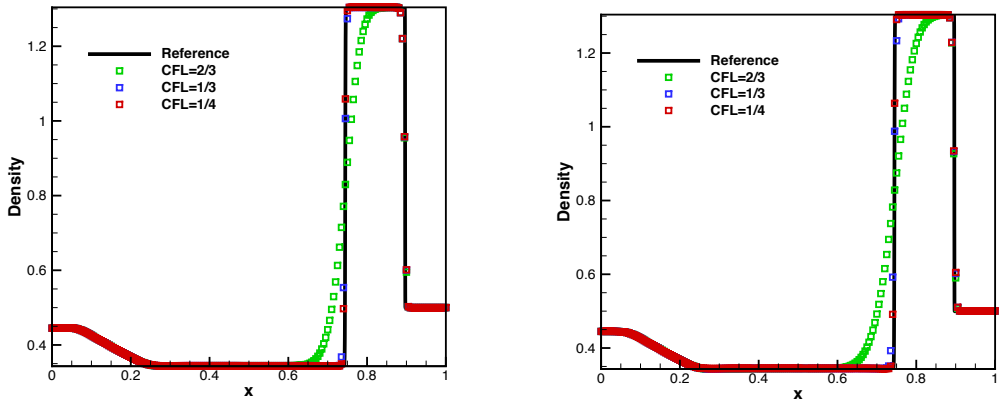

Figure 25: Lax problem with RK3. (left) using SAH (supremum); (right) using THINC (supremum).

Again, all solutions are in line with the reference solution for the right and left rarefaction waves. All schemes also present numerical dissipations at the contact discontinuity for $\nu>0.5$. However, when $\nu<0.5$, the SAH (supremum) and THINC (supremum) schemes show good agreement, that is, a sharp contact discontinuity, with the reference solution. The self-adjusting steepness-based schemes also show this property for smaller CFL numbers.

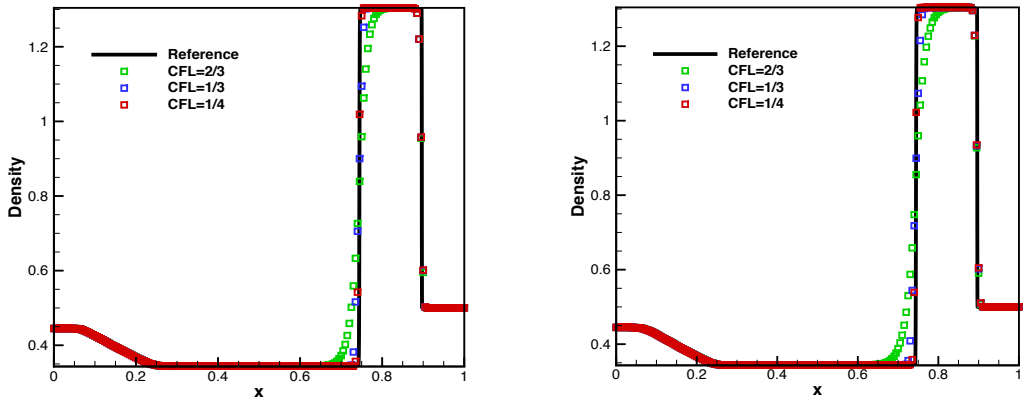

Figure 26: Lax problem with RK3. (left) using SAS (SAH); (right) using SAS (THINC).

*5.3.3 Two-blast wave problem*

We also consider a two-blast-wave interaction problem proposed by Woodward and Colella [32]. The initial condition is

$$(\rho, u, p) = \begin{cases} (1,0,1000), & \text{if } 0 \leq x \leq 0.1 \\ (1,0,0.01), & \text{if } 0.1 \leq x \leq 0.9 \\ (1,0,100), & \text{if } 0.9 \leq x \leq 1. \end{cases}$$

and $t = 0.038$. The reflective boundary conditions are applied at $x = 0$ and $x = 1$. We examine the numerical solutions on a 401-point grid. Figs.27-28 depict the density distributions computed using different numerical schemes for different CFL numbers.

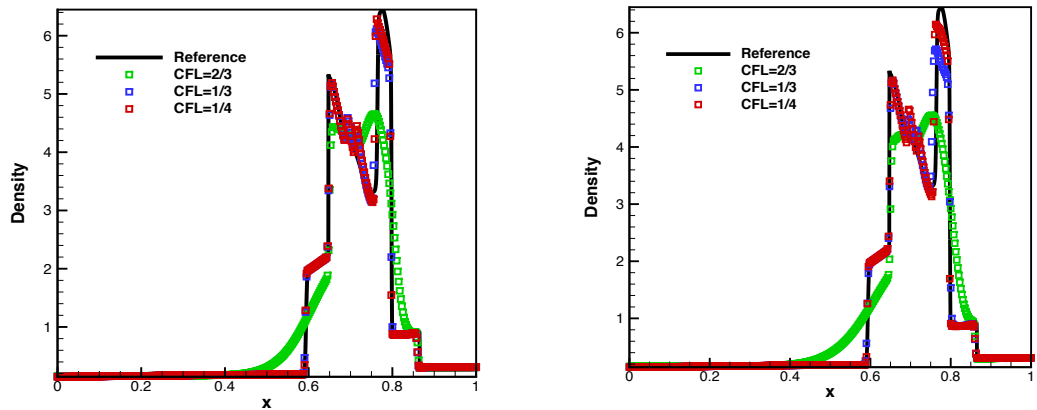

Figure 27: Two blast wave problem with RK3. (left) using SAH (supremum); (right) using THINC (supremum).

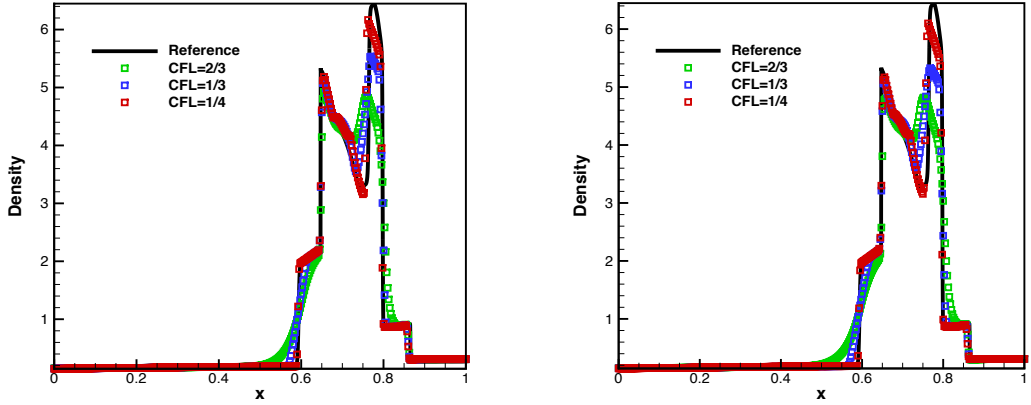

Figure 28: Two blast wave problem with RK3. (left) by SAS (SAH); (right) using SAS (THINC).

Again, all solutions agree well with the reference solution for the shocks and rarefaction waves. All schemes also present numerical dissipations at the contact discontinuity for $\nu > 0.5$. When $\nu < 0.5$, the supremum and supremum schemes exhibit a sharp contact discontinuity. The self-adjusting steepness-based schemes also show such properties for smaller CFL numbers.

However, the problem now occurs in the region, where the two initial shocks interact with each other. The SAH (supremum) and THINC (supremum) show some oscillations in this region. This is because these schemes are not applicable to smooth structures. In contrast, the self-adjusting steepness-based schemes show a very good agreement with the reference in this region. This is because of their greater numerical dissipation as enabling them to preserve contact discontinuity for smaller CFL numbers.

*5.3.4 Sedov problem*

The Sedov blast wave test [33] involves low density and low pressure. The initial conditions are:

$$(\rho, u, p) = \begin{cases} (1, 0, (\gamma - 1) \times 10^{-12}), & \text{if } x = 0, \\ \left(1, 0, \dfrac{\gamma - 1}{dx} \times 3.2 \times 10^6\right), & \text{otherwise.} \end{cases}$$

Such a high-energy deposition generates strong left and right moving shocks followed by an exponential decay of density and pressure. These exponential decays lead to an almost near-vacuum region in the center of the domain. Here we consider the computational domain of $[-2,2]$ on an 801 point grid. The zero-gradient boundary condition is also applied at $x = -2$ and $x = 2$ and $t = 1 \times 10^{-3}$. Figs.29-30 depict the density distributions computed using different numerical schemes for different CFL numbers.

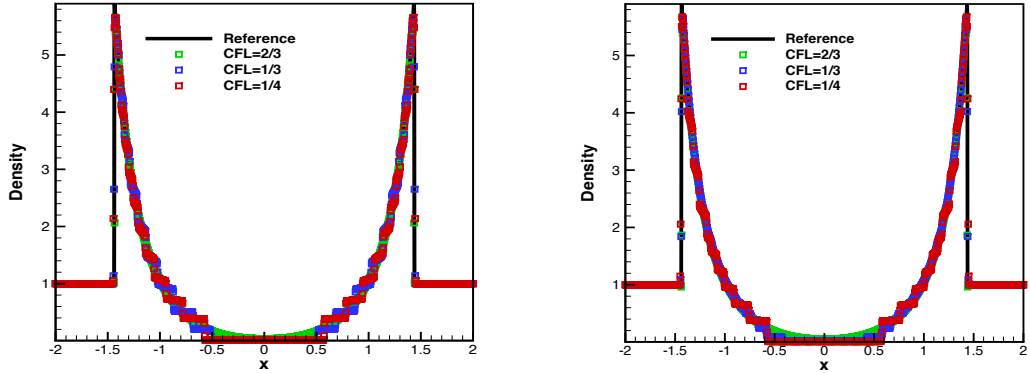

Figure 29: Sedov problem with RK3. (left) using SAH (supremum); (right) using THINC (supremum).

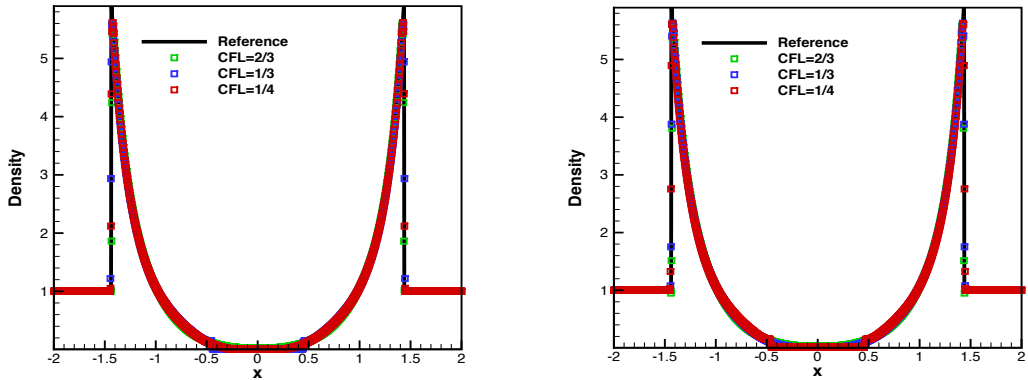

Figure 30: Sedov problem with RK3. (left) using SAS (SAH); (right) using SAS (THINC).

It is also seen that all solutions agree well with the reference solution for the shocks. As there is no contact discontinuity, the supremum (SAH) and supremum (THINC) schemes result in over-compression for the rarefaction waves. Decreasing the CFL number, the over-compression is amplified. However, this over-compression is reduced for self-adjusting steepness-based schemes. Finally, this test indicates that the self-adjusting steepness-based schemes are capable of capturing vacuum.

*5.3.5 123 problem*

The exact solution of 123 problem [33] contains a vacuum. The initial conditions are

$$(\rho, u, p) = \begin{cases} (7, -1, 0.2), & \text{if } -1 < x \leq 0, \\ (7, 1, 0.2), & \text{if } 0 \leq x \leq 1. \end{cases}$$

The computational domain is $[-1,1]$ on a 401 point grid. We use the zero-gradient boundary condition at $x = -1$ and $x = 1$, and $t = 0.6$. Figs.31-32 depict the computed density distributions by different numerical schemes for different CFL numbers.

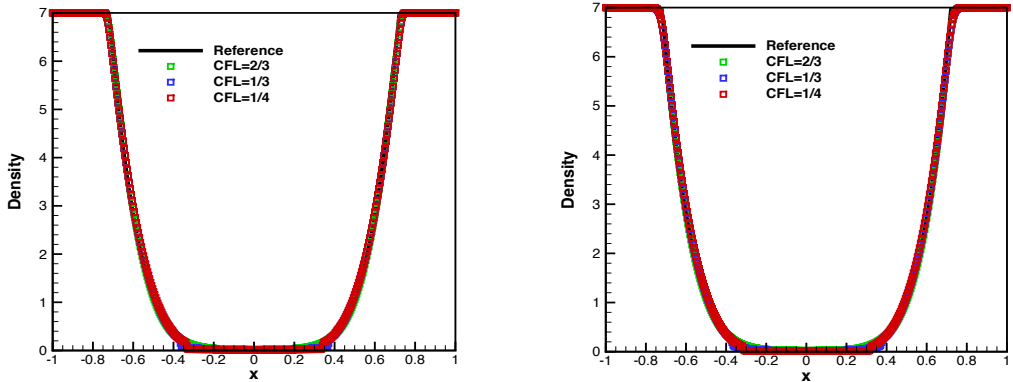

Figure 31: 123 problem with RK3. (left) using SAH (supremum); (right) using THINC (supremum).

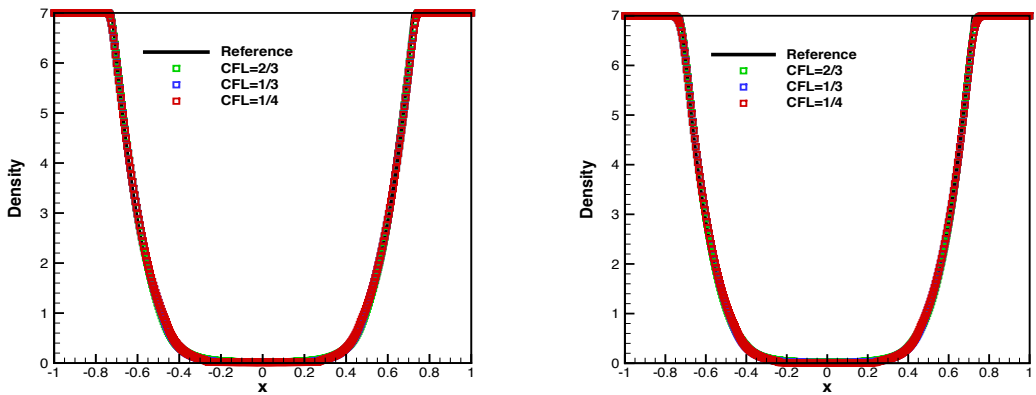

Figure 32: 123 problem with RK3. (left) by SAS (SAH); (right) using SAS (THINC).

The solution to this problem is two rarefaction waves and a vacuum zone. The SAH (supremum) and THINC (supremum) schemes show their capacity for compression at the boundary of the vacuum zone $|x| \simeq 0.35$. The over-compression does not occur in self-adjusting steepness-based schemes. This test also shows that the self-adjusting steepness-based schemes are capable of computing continuous regions and even vacuum regions.

## 5.4 Two-dimensional Euler equations

In this section, we solve the two-dimensional Euler equations. The numerical solutions are computed only by the self-adjusting steepness-based schemes. The third-order TVD Runge-Kutta scheme [14] is also used for time discretization.

### 5.4.1 Double Mach reflection

We consider the problem from Woodward and Colella [32] on the double Mach reflection of a strong shock. A Mach 10 shock in the air is reflected from the wall with an incident angle of $60°$. For this case, the initial condition is

$$(\rho, u, v, p) = \begin{cases} (1.4, 0, 0, 1), & \text{if } y \leq 1.732(x - 0.1667), \\ (8, 7.145, -4.125, 116.83333), & \text{else,} \end{cases}$$

and $t = 0.2$. The computational domain is $[0,4] \times [0,1]$ with $961 \times 241$ grid points and the reflecting wall lays at the bottom of the computational domain for $1/6 \leq x \leq 4$. Initially, a right-moving Mach= 10 shock wave is located at $(x, y) = (1/6, 0)$ and forms a $60°$ with the $x$-axis. The right unshocked air has a density of 1.4 and a pressure of 1. For the boundary, we impose an exact post-shock condition for the area $x \in [0, 1/6]$. Inflow and outflow boundaries are used for the left and right boundaries, respectively. The values at the top boundary are also set to describe the exact motion of a Mach 10 shock.

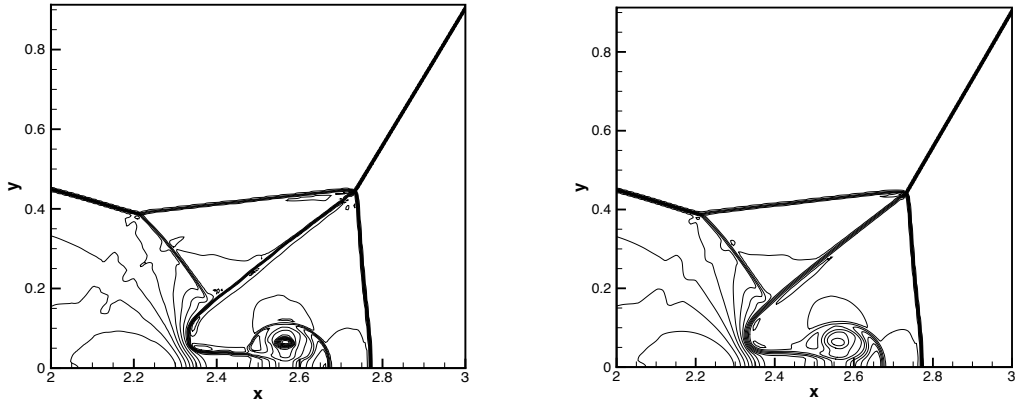

Figure 33: Density distribution of double Mach reflection with 30 contours between 2 and 22 obtained using SAS (SAH) scheme. (left) ν=1/4; (right) ν=1/3.

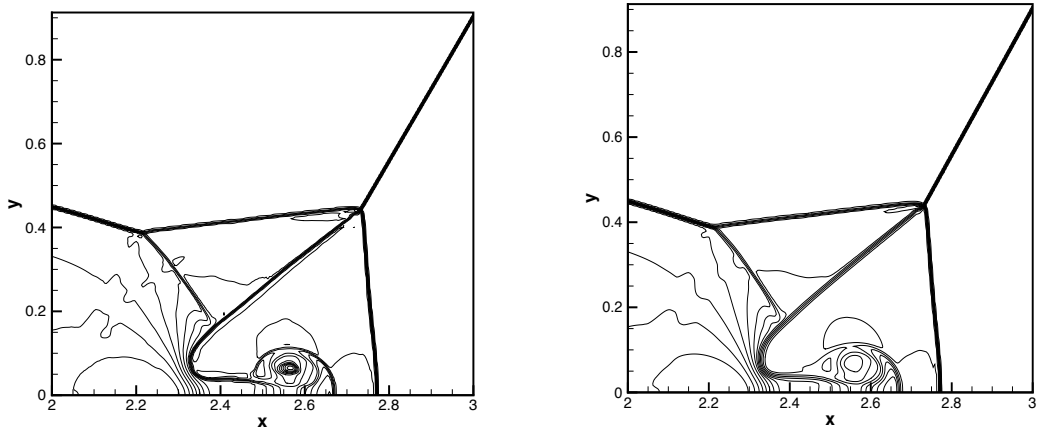

Figure 34: Density distribution of double Mach reflection with 30 contours between 2 and 22 obtained using SAS (THINC) scheme. (left) ν=1/4; (right) ν=1/3.

Fig.33 and Fig.34 depict the density contours of the solution computed using the SAS (SAH/THINC) schemes, respectively, under $\nu = 1/4$ and $\nu = 1/3$. The solutions under $\nu = 1/3$ show a contact discontinuity of equal quality to that of the WENO5 scheme. The contact discontinuities in the solutions under $\nu = 1/4$ are thinner than those under $\nu = 1/3$. This phenomenon agrees with the theory that the smaller the CFL number, the more anti-diffusions there are, which leads to a sharp contact discontinuity.

Moreover, under both CFL numbers, there were several numerical oscillations. This again confirms the importance of the $\beta$-limiting process.

*5.4.2 Richtmyer-Meshkov instability*

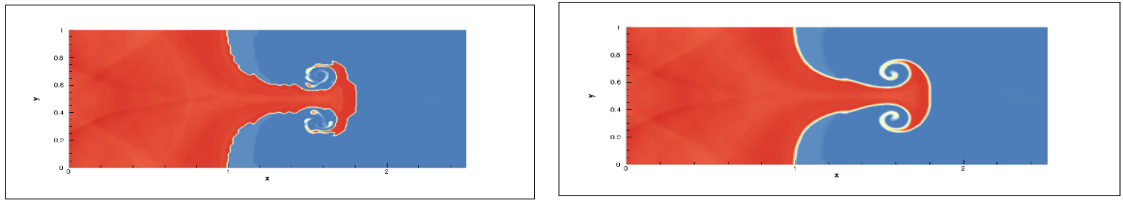

Figure 35: Density distribution of RM instability with 30 contours between 3 and 13 obtained using the SAS (SAH) scheme. (left) ν=1/4; (right) ν=1/3.

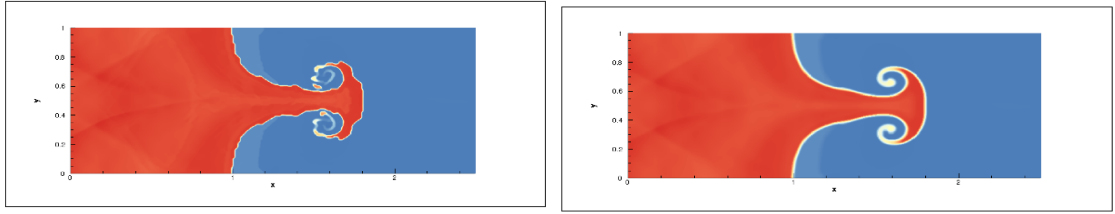

Figure 36: Density distribution of RM instability with 30 contours between 3 and 13 obtained using the SAS (THINC) scheme. (left) ν=1/4; (right) ν=1/3.

Richtmyer-Meshkov instability is a major phenomenon that shuts down the ignition of inertial confinement fusion. Fluid instability occurs if a shock wave interacts with a perturbed interface between a light fluid and a heavy fluid. The time evolution of the Richtmyer-Meshkov instability depends on the interface capturing technique, i.e., a sharp interface provides more details regarding the fluid structures.

Richtmyer-Meshkov instability is also examined in this study. Initially, a material interface of $x = 3.9 + 0.1\sin(2\pi(y + 3/4))$ is placed in the computation domain of $[0,5] \times [0,1]$ with $1281 \times 257$ grid points. At the left of the interface, the air is set to $\rho = 5.805$. At the right of the interface, the air is set to $\rho = 1.153$. A left-moving Mach= 2.4 shock wave is also located at $x = 4.2$. Across the interface, the pressure is set to $p = 1/\gamma$, where for air $\gamma = 1.4$. Periodic boundary conditions are applied at the top and bottom. The initial conditions are imposed on the left and right sides, and a result of $t = 5$ is obtained.

The Fig.35 and Fig.36 show the density contours of the solution computed by the SAS (SAH/THINC) schemes, respectively, for $\nu = 1/4$ and $\nu = 1/3$. These results confirm the conclusions stated previously.

## Conclusion

In the self-adjusting steepness (SAS)-based schemes, different values of the steepness parameter $\beta$ in the steepness-adjustable limiters result in different behaviors. These behaviors include a high-order property with the exact value of the steepness parameter theoretically determined or an anti-diffusive/compression property with a larger steepness parameter. In this study, we demonstrate that the limit of the steepness-adjustable limiter is the ultrabee limiter. We also theoretically derive the analytical expressions for the supremum of the steepness parameter for two steepness-adjustable limiters. Based on these we then propose novel supremum-determined SAS schemes. These schemes are further extended to solve Euler equations which are applied only to linearly degenerate fields. The numerical results validate our analysis and the final schemes are efficient in capturing contact discontinuities. It is also shown that using our proposed schemes the anti-diffusion level is controlled and does not cause numerical oscillations, which usually lead to nonphysical small structures, such as small vortices.

In the future, further investigations are required for the supremum-determined self-adjusting steepness-based schemes. Examples are on their multi-dimensional performance. We observed a slight asymmetry in the two-dimensional Richtmyer-Meshkov tests. This asymmetry might be solved using a multidimensional limiting process. Furthermore, it would be interesting to extend the present scheme to higher orders.

## Acknowledgement

We acknowledge financial support from the President Foundation of CAEP under Grant No. YZJJLX2018012 and national key project under grant No. GJXM92579.